\documentclass{desyproc}

\def\lesssim{\ \hbox{\raise 2pt \hbox{$<$} \kern -13pt
                     \lower 3pt \hbox{$\sim$}}\ }
\def\greatersim{\ \hbox{\raise 2pt \hbox{$>$} \kern -13pt
                     \lower 3pt \hbox{$\sim$}}\ }
\def\kt{\ensuremath{k_\perp}}

\def\cascade{{\sc Cascade}}
\def\ldcmc{{\sc Ldcmc}}

\begin{document}

\title{   \hspace*{0.4 cm}  TMD PDFs: a Monte Carlo implementation 
 \hspace*{3.1 cm}  for  the sea quark distribution
}

\author{    \hspace*{3.95 cm} 
F.~Hautmann$^1$,   M.~Hentschinski$^2$ and H.~Jung$^{3,4}$
%
\\ 
 \hspace*{1.5 cm} 
$^1${\slshape Department of  Theoretical Physics, University of Oxford, 
Oxford OX1 3NP}\\   
 \hspace*{0.1 cm} 
$^2${\slshape Instituto de F{\' i}sica Te{\' o}rica  
UAM/CSIC,  Universidad Aut{\' o}noma de Madrid, 
E-28049 Madrid}\\ 
 \hspace*{3.4 cm} 
$^3${\slshape  Deutsches Elektronen Synchrotron, D-22603 Hamburg}  \\ 
 \hspace*{3.8 cm} 
$^4${\slshape  CERN, Physics Department, CH-1211 Geneva 23} 
}

\maketitle

\begin{abstract} 
This article   gives an  
 introduction to  transverse momentum dependent (TMD)  
parton distribution functions and their use in  shower Monte Carlo event generators 
for high-energy hadron collisions, and describes  recent progress in the treatment 
of   sea quark  effects within a TMD  parton-shower  framework.
\vskip 0.2 cm   \hspace*{2.55 cm} 
Presented at the Conference {\em Photon 2011}, Spa, May 2011 
\vskip -8.8cm
\hspace*{9.0 cm} {  DESY 12-081 } \\ 
\hspace*{9.0 cm} {  IFT-UAM/CSIC-12-48 } \\ 
\hspace*{9.0 cm} {  OUTP-11-55-P }
\vskip 8.2cm 
\end{abstract}

\section{Introduction}

While  QCD factorization methods are  well established 
in the  case of  scattering observables involving  
a single large mass  scale~\cite{css04},   
the treatment  of processes with multiple mass scales is  
rather more  subtle.  In multi-scale processes, 
generalized factorization formulas  are 
 required~\cite{jcc-book}     if  one is to control  perturbative  large 
logarithms to higher orders of perturbation theory  and 
  to  describe appropriately nonperturbative physics in the initial 
   and final states of the collision. 
Such generalized  factorization  formulas typically  involve 
transverse-momentum dependent  (TMD), or  ``unintegrated",     
parton distribution  and parton decay functions~\cite{jcc-book,mert-rog}.

A full treatment of factorization at   unintegrated 
 level, valid uniformly over the whole 
space, is yet to be achieved~\cite{muld-rog}.  
Results however exist  which apply 
 in specific phase space regions. One such case is 
given by  QCD in the high-energy, or 
small-$x$,  limit~\cite{lip97}, in which 
 factorization of    TMD   gluon distributions   holds~\cite{hef},    in 
 correspondence to  perturbatively-resummed coefficient functions.  
Results based on small-$x$ TMD distributions  have  for instance been used 
 in  analyses of  inclusive observables such as deeply inelastic 
structure functions (see~\cite{abf09} for a review) and   in 
 Monte Carlo simulations of the  
exclusive  structure of   final states   in  hadronic 
collisions  (see~\cite{hj_rec}).

Many  aspects  of     the experimental program at the 
Large Hadron Collider (LHC)  
depend on  the analysis of  processes 
containing multiple hard scales, and will   be influenced  by improved 
  formulations  of factorization  in QCD at unintegrated level. 
In this article we give a concise overview of   
recent  progress   on  TMD  distributions  
and  their   use for simulations of final states  in hadronic collisions 
    by   shower Monte  Carlo event generators.   
Most of  TMD 
computational tools have so far been developed   within 
a quenched approximation in which only gluon and valence quark 
effects  are taken    into   account   at TMD level.  We describe results of 
recent work~\cite{hent11,hent-dis}  to   go beyond this 
approximation by   including   sea quark effects,  and we 
present  first  numerical  applications  to  forward  $Z$-boson  production. 

The  article  is organized as follows. In Sec.~2 we  recall 
 operator definitions  for parton distribution functions (PDFs) and 
basic  issues   associated with their   TMD generalization. 
In  Sec.~3  we  consider  parton branching methods and   the 
role of  unintegrated    distributions in shower Monte Carlo 
generators.  In Sec.~4 we discuss 
the approach~\cite{hent11}   to incorporate 
effects from  quark emission and  flavor-singlet   sea-quark 
contributions   in the  framework of 
transverse-momentum dependent parton showers.   We summarize in Sec.~5.

\section{TMD parton distribution functions}

In this section we introduce the   parton correlation functions used to define 
parton distributions.  In Subsec.~2.1 we  consider 
 operator matrix elements  for TMD distributions, and  discuss   current  open   
 issues,  including lightcone divergences and factorization breaking effects.  
In Subsec.~3.2  we  focus on  the case of  small $x$, related factorization 
results, and introduce applications which are the subject 
of  the  sections that follow.

\subsection{Operator  matrix  elements}

\begin{figure}[htb]
\vspace{40mm}
\includegraphics{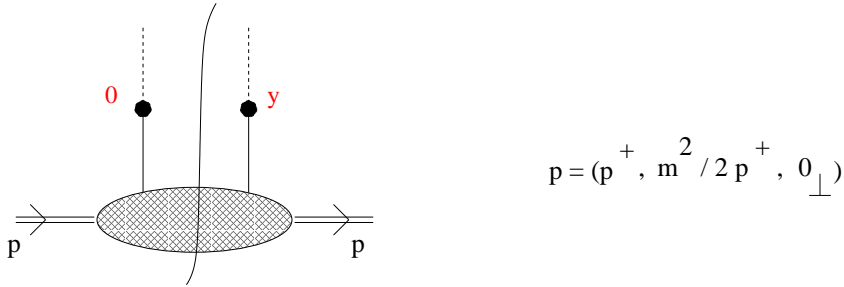}
\caption{\it \small   Correlator of two quark  fields  at  distance $y$.} 
\label{fig:tmdop}
\end{figure}

The relevance of consistent operator definitions for 
parton k$_\perp$ distributions  was emphasized 
long ago   in the context of 
 Sudakov processes~\cite{collsud}, 
  jet physics~\cite{cs81}, 
exclusive production~\cite{brodlep}, 
spin physics~\cite{mulders}. 
The  approach commonly used  to ensure gauge invariance 
   is to  generalize     
 the   coordinate-space     matrix elements  that define
     ordinary parton distribution 
 functions (pdfs)~\cite{cs81-82}  
to the case  of    field   operators  at non-lightcone distances. 
For instance,   for  the quark  distribution 
 one has   (Fig.~\ref{fig:tmdop})   
\begin{equation}
\label{coomatrel}
  {\widetilde f} ( y  ) =
  \langle P |  {\overline \psi} (y  )
  V_y^\dagger ( n ) \gamma^+ V_0 ( n )
 \psi ( 0  ) |
  P \rangle  \hspace*{0.3 cm}  .   
\end{equation}
Here  $\psi$ are  the quark fields evaluated at distance 
$y = ( 0 , y^- , y_\perp  )$,     
      where  $y_\perp$ is in general nonzero, 
 and   $V$ 
 are  eikonal-line operators in direction $n$,  
\begin{equation}
\label{defofV}
V_y ( n ) = {\cal P} \exp \left(  i g_s \int_0^\infty d \tau \
n^\mu A_\mu (y + \tau \ n) \right)
    \hspace*{0.3 cm} ,    
\end{equation}
which we require to   make the matrix element gauge-invariant. 
 The unintegrated, or TMD,  quark distribution is obtained from the double 
 Fourier transform in  $y^-$ and $y_\perp$ of ${\widetilde f}$.

While  Eq.~(\ref{coomatrel})   works  at tree level     (including an 
extra gauge link at infinity in the case of physical gauge~\cite{beli},  see 
also~\cite{bacch-diehl}),   
going beyond tree level   requires  treating 
 lightcone  singularities~\cite{collsud,brodsky01},   associated with the 
$x \to 1$ endpoint.  
The   singularity structure at $x \to 1$  is different  
 than for 
ordinary (integrated) distributions, giving divergences  
 even in dimensional regularization with 
an infrared cut-off~\cite{fhfeb07}.  
 The singularities can be understood 
in terms of gauge-invariant eikonal-line matrix elements~\cite{fhfeb07}  
and  related to cusp anomalous 
dimensions~\cite{korchangle,korchangle_marche,chered}.  
 
This  can be   analyzed  explicitly   
at one   loop.   Expansion in powers of $y^2$ of the 
  coordinate-space matrix 
element  (\ref{coomatrel}) at this order  
gives~\cite{fhfeb07,fhdistalk} 
 \begin{eqnarray}
\label{EbmEaexpand}
 {\widetilde f}_{1} (y)  &=&   
{{ \alpha_s C_F     } \over { \pi  } } 
\ p^+
\int_0^1 dv  \ { v \over { 1 - v }} \ 
\left\{ 
\left[ e^{ i p \cdot y v}  - e^{ i p \cdot y } \right] 
\ \Gamma ( 2 - { d \over 2} ) \ 
( { {4 \pi \mu^2} \over \rho^2} )^{2-d/2}
\right. 
\nonumber\\
 &+& \left.
e^{ i p \cdot y v} \ \pi^{2-d/2} \ 
\Gamma (  { d \over 2} - 2 ) \ 
(- y^2  \mu^2)^{2 - d/2}
+ \cdots 
\right\} \hspace*{0.2 cm}  , 
\end{eqnarray}
where $d$ is the number of space-time     
    dimensions,   $\mu$ is the  dimensional-regularization scale  
  and $\rho$ is the  infrared  mass regulator.  
  The first term in the right hand side of Eq.~(\ref{EbmEaexpand}) 
  corresponds to the case of  ordinary pdfs. 
The lightcone singularity $v \to 1$,  
corresponding  to the exclusive  boundary $x = 1$,   
  cancels in this term,  but it    
 is present, even at $d \neq 4$ and finite 
$\rho$,   in subsequent terms. 

These  endpoint singularities  come  from 
gluon emission at large rapidity.  
They imply that,   using 
the matrix element (\ref{coomatrel}),     in momentum space 
the $1/ (1-x)$ factors from 
 real emission probabilities do not  in general  combine with 
 virtual corrections to give  $1/ (1-x)_+$             
 distributions, but leave uncancelled divergences at fixed 
 k$_\perp$.   It is only after supplying the above matrix element 
with a regularization prescription that the distribution  is 
well-defined. 

A  possible regularization  method for the endpoint is 
by  cut-off, implemented by taking the eikonal line $n$  in 
Eq.~(\ref{defofV}) to be 
non-lightlike~\cite{collsud,korch89}, 
combined with 
evolution equations in the cut-off 
parameter $\eta = (p \cdot n)^2 / n^2$~\cite{cs81,korchangle}.
Then the  cut-off 
in $x$ at fixed  $k_\perp$  is of order 
$ 1 - x   \greatersim k_\perp / \sqrt{4 \eta}$.  
 Monte Carlo event  generators that  make use of 
 unintegrated pdfs      also  implement  a  cut-off.  We  consider  such   
 applications  in Sec.~3.

\begin{figure}[htb]
\vspace{33mm}
\includegraphics{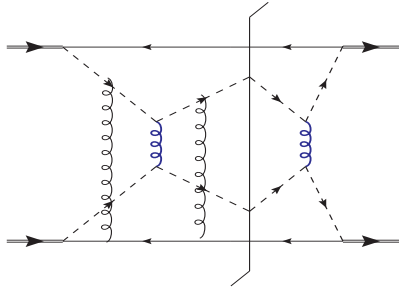}
\caption{\it \small  Soft gluon exchange with   
spectator partons.} 
\label{fig:spect}
\end{figure}

An alternative method is provided by the 
   subtractive method~\cite{jccfh00,jccfh01}, in which  the 
direction $n$   in 
Eq.~(\ref{defofV})   is kept  lightlike  but the divergences 
are canceled by multiplicative, gauge-invariant counterterms 
given by   vacuum expectation  
values of eikonal operators.    This  method 
leads to well-prescribed 
counterterms~\cite{fhfeb07}  for the 
transverse momentum dependent splitting  probabilities, which can be  
viewed as   generalizing the   plus-distribution regularization 
 for  $ k_\perp \neq 0$.   
 On  one hand, 
 this approach  has been   used to   relate  the endpoint behavior   at fixed 
k$_\perp$  with  the  cusp anomalous  dimension~\cite{chered} 
and investigate  the role of  the Mandelstam-Leibbrandt prescription 
in lightcone gauges.  On the other hand,  it 
can  likely  be more useful   than the cut-off to  untangle    issues of 
factorization   and non-universality~\cite{jcc-book,muld-rog,nonfac07,li-11}   
which arise beyond leading order, and investigate the relationship between 
  evolution equations in rapidity and 
  in virtuality~\cite{cs81,korchangle,ceccopie}. 
 Infrared   subtractions     analogous to 
 those    in the  method~\cite{jccfh00,jccfh01}     
  are  also   discussed  in the context of the 
soft-collinear effective theory~\cite{manohstew}    
(under the form, however,  of counterterms   that  are not automatically gauge-invariant)   
 in the case of  the Sudakov form factor~\cite{rean1}    and 
  of   initial-state  beam functions~\cite{beam-f}   describing the incoming jet  
  (see also~\cite{idi-sci}). 

As noted earlier, 
 a  full treatment of factorization at  the level  of TMD  pdfs 
  is yet to be achieved~\cite{jcc-book,muld-rog}.  
In the   hadroproduction of nearly  back-to-back  
 hadrons,     factorization is broken~\cite{muld-rog}    by     soft gluons exchanged 
between  subgraphs in different collinear  directions   (Fig.~\ref{fig:spect}). 
   (See also the  analyses~\cite{becher-neu,man-petrie} for   the Drell-Yan case). 
The underlying 
dynamics is that of  non-abelian 
Coulomb phase~\cite{aybat},    involving   interactions with spectator 
partons~\cite{boerbrod} (which 
 were treated      long  ago in~\cite{css04}   limited  to   
 the  case of   fully inclusive Drell-Yan).  
 It was noted in~\cite{hj_rec} 
 that the factorization-breaking  contributions~\cite{nonfac07,muld-rog} 
 are Coulomb/radiative mixing terms 
 related to the contributions~\cite{manch1}  responsible for the appearance of 
 super-leading logarithms  in  di-jet cross sections with a gap in rapidity.  

The   issue of    factorization      depends  on 
 developing a systematic treatment,   capable of  handling   
 overlapping divergences in infrared regions    
for complex observables that   
 involve color charges    in both initial and final states.  
 This 
    motivates  
 the  subtraction   techniques 
quoted  above, in the  version~\cite{jccfh00,jccfh01} or in the 
SCET  version~\cite{manohstew,rean1,beam-f,idi-sci,becher-neu,man-petrie}.  

Factorization   involves in general,  
besides TMD pdfs,   a nonperturbative soft factor, 
also characterized  in terms of operator matrix elements, 
see~\cite{jccfh00}. The first paper in 
Ref.~\cite{jccfh01}  analyzed the  possibility  of 
 reabsorbing  the soft factor into a redefinition of the TMD pdfs (and analogous 
final-state  fragmentation functions), 
 using  an explicit  construction at one loop   in the case of 
 initial-state and final-state parton showers in  DIS.   
 More recently,  an explicit 
   procedure to reabsorb  the soft factor into 
 redefined TMD pdfs 
 has been presented in Ref.~\cite{jcc-book}  
 in the context of Drell-Yan.   An analogous redefinition is  
 implied by  the  Drell-Yan analyses in~\cite{becher-neu,man-petrie}. 
  It  remains to be seen   how generally this can be 
  done. This will influence  future    programs  of  phenomenological  
  determinations  of TMD pdfs, see~\cite{mert-rog}.

\subsection{TMD formulation at small $x$ }

The case of back-to-back  di-hadron or di-jet 
 hadroproduction~\cite{muld-rog,nonfac07}   
 illustrates that, due to the difficulty in disentangling   soft and collinear gluon 
correlations between initial and final states, a   general 
 TMD factorization formula is still lacking. 
In the case of small $x$, however, a TMD factorization result 
holds~\cite{hef} owing to the 
dominance of single gluon helicity at high energy.  In this case, a TMD gluon 
distribution can be defined gauge-invariantly from the high-energy pole in physical 
cross sections.  See~\cite{mue11,avsar11,qiuetal11,doming11} for  recent  
discussions  of unintegrated pdfs  based on   small $x$.  

\begin{figure}[htb]
\vspace{38mm}
\includegraphics{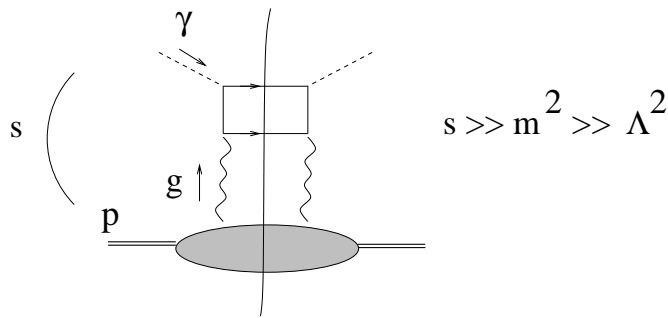}
\caption{\it \small 
TMD factorization at high energy for heavy quark photoproduction.} 
\label{fig:disx-acta}  
\end{figure}

The main reason why   such   a  definition for TMD 
pdfs can be constructed in the high-energy limit   
is that   one can relate directly (up to 
perturbative corrections) the cross section for a {\em physical}  
process, e.g. photoproduction of a heavy-quark pair~\cite{hef90}, 
to an   unintegrated,  transverse momentum dependent 
 gluon distribution (Fig.~\ref{fig:disx-acta}).  
 This is   quite like what one does for  deeply inelastic scattering, in 
 the conventional 
parton picture,   in terms of ordinary (integrated) parton distributions. 
On the other hand, the difficulties 
in defining a TMD distribution   in the general case, over the whole phase space,  can 
 be associated  with the fact that 
it is not  obvious how to determine one such 
 relationship  for general kinematics. 

The evolution equations obeyed by TMD distributions defined from the small-$x$  
limit are  evolution equations in energy~\cite{lip97}, with   corrections  
 down by  logarithms of energy rather than  powers of momentum transfer. 
 Using this evolution,    
 the high-energy factorization~\cite{hef}  can be related, order-by-order in 
 $\alpha_s$~\cite{ch94}, with  the renormalization-group 
   factorization~\cite{css04,cs81-82,CFP}. This   allows one 
 to describe in this framework 
 the ultraviolet region of arbitrarily high k$_\perp$,    
and in particular re-obtain the structure of QCD 
logarithmic scaling violations, see 
e.g.~\cite{abf09,xlog}. 

  The above observation 
justifies the use of this approach for  hard production physics.       
 In particular, it is the basis for  using  
   Monte Carlo implementations   
 of  transverse momentum dependent  parton showers 
 (see e.g.~\cite{hj_rec,acta09} and references therein) 
  to treat multi-scale hard processes at the LHC. 
Parton-shower   applications   are discussed 
in Sec.~3.  Such  applications  have so far focused on TMD gluon 
distributions.  Ongoing work on  the  generalizations    needed to include 
quark channels is discussed in Sec.~4.

 Let us finally recall that  
  extensions  of  the   factorization 
  results  above   are  required if 
one is to take into account     the nonlinear effects that are  expected  to arise 
in the small $x$  region  from     high  parton densities.       Recent  
work in this direction, focusing on multiple scattering effects in  
 dense targets and  nuclei,    may be found in~\cite{doming11,xiao,xiao12}.  
 In this respect, we 
note that 
 techniques such as those in~\cite{s-channel}   have been proposed 
to  incorporate 
 the treatment of     multiple-gluon rescattering graphs  at small $x$   
starting from  the operator matrix 
elements~\cite{collsud,cs81-82}  for parton distributions. 
They may   thus  be  helpful    for   extensions  to  the high density region 
that are aimed at   retaining   accuracy  also in the 
treatment of  contributions from high p$_{\rm{T}}$ processes.

\section{Parton-branching  applications}

In this  section we move to   applications 
of the  TMD formalism  to parton branching methods. 
The main role   of TMD  splitting functions and distributions 
 in this  context is that they serve  to 
 take into account coherence effects of multiple gluon radiation 
for  small  longitudinal momentum fractions $x$ 
   in the initial  state  parton cascade. 
In Subsec.~3.1 we  briefly 
recall   the motivation for     angular-ordered parton showers, 
and in Subsec.~3.2  we  discuss the 
treatment of gluon coherence at small $x$ and 
corrections to angular ordering.

\subsection{Collinear showering and soft gluon coherence} 

\begin{figure}[htb]
\vspace{40mm}
\includegraphics{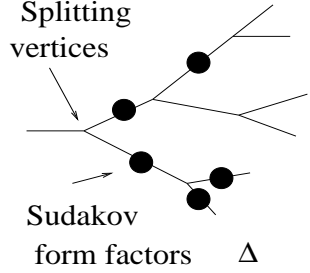}
\caption{\it   \small  
Parton  branching in terms of splitting probabilities and form factors.}
\label{fig:pshower}
\end{figure}

Branching algorithms in  standard 
shower Monte Carlo generators~\cite{mc_lectures}   are 
based  on collinear evolution 
of the jets, both  time-like and space-like,  
developing from  the hard event. The branching probability  is given in terms 
of splitting functions $P$  and form 
 factors $\Delta$ (Fig.~\ref{fig:pshower}) as  
\begin{equation}
\label{psbasic}
 d {\cal P} = 
 \int { { d q^2} \over q^2} \  
  \int { {dz} } \  \alpha_S   (q ^2)   \  { P} ( z) \   
 \Delta    (q^2 ,   q_0^2)     \;\; . 
\end{equation}
The theoretical basis for the branching approach 
  is the factorizability  of universal 
splitting functions in QCD cross sections in the  
collinear limit~\cite{css04,eswbook}, which  
justifies   the probabilistic picture. 

\begin{figure}[htb]
\vspace{50mm}
\includegraphics{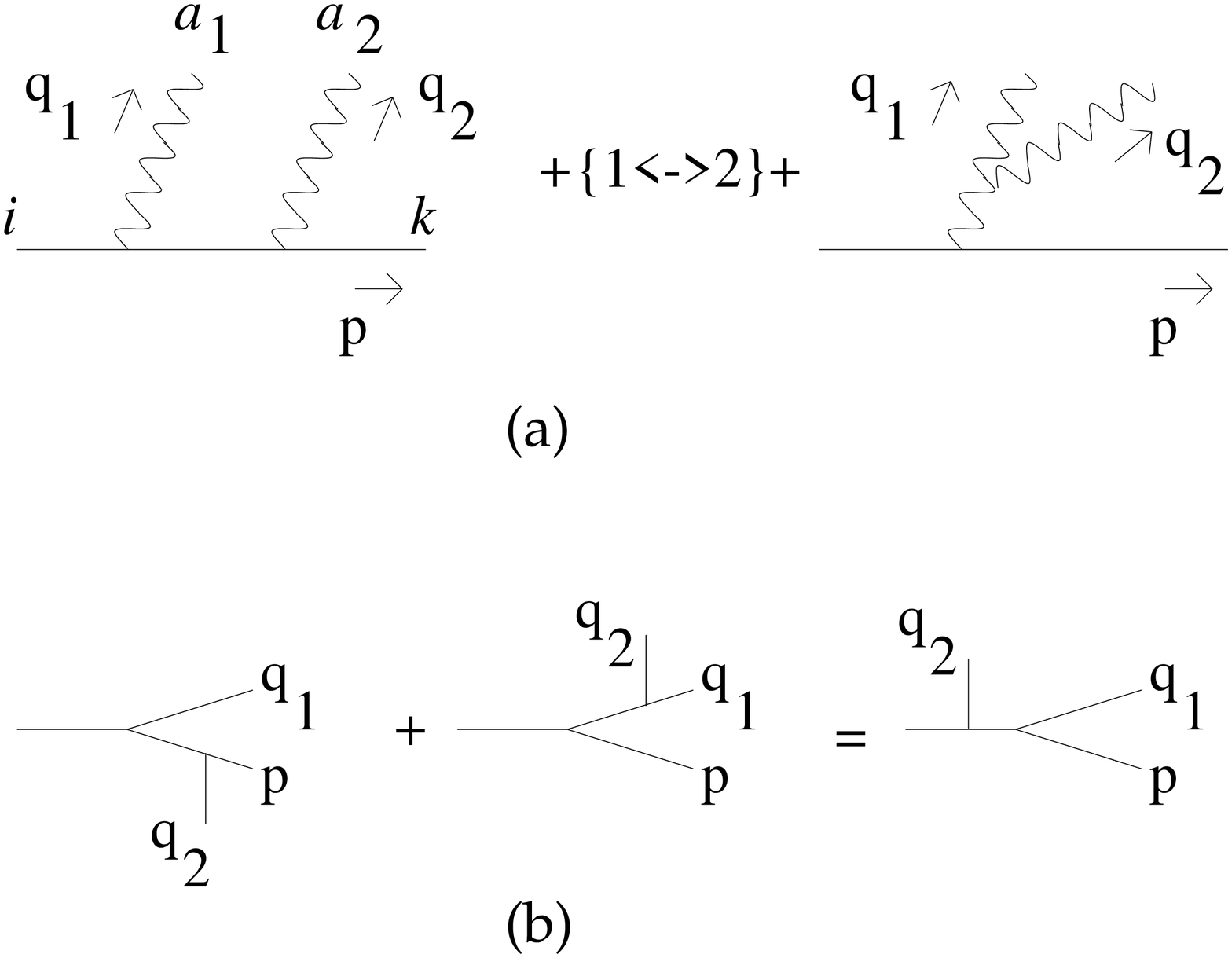}
\caption{\it   \small   (a) Two gluon emission from a fast quark; 
 (b) coherence of soft gluon emission at large angle.}
\label{fig:colcoh} 
\end{figure}

Besides small-angle, incoherent  parton emission,  many of the current shower 
generators  also take into account   further 
radiative  contributions from  emission of  soft gluons, which 
are essential for realistic  phenomenology~\cite{mc_lectures}.  
To incorporate  these  in a probabilistic framework,    one   
appeals  to properties of coherence  of  color 
radiation~\cite{eswbook,bcm,dokrev}.   
 Soft-gluon  emission amplitudes factorize 
in terms of  eikonal  currents~\cite{griblow,jctaylor}   
\begin{equation} 
\label{eikqcd}
    {\bf J}_\mu^a = \sum_{i=1}^{n}  {\bf Q}_i^a \  
    {p_{i \mu} \over { p_i \cdot q} } \;\;\; , 
\end{equation}
where $p_i$ are the emitters'  momenta, $q$ is the soft momentum, and 
the   color charge operators 
${\bf Q}^a_i$ are   
associated with the emission of gluon $a$ from parton $i$.  
In general, 
interferences are  expected to 
contribute to the radiative terms  relating the $(n+1)$-parton process to the 
$n$-parton process. 
Nevertheless,  a probabilistic 
branching-like picture can be 
recovered~\cite{marchweb80s,dok88,mc89}  by exploiting 
 soft-gluon  coherence. This   is illustrated  in 
  Fig.~\ref{fig:colcoh}~\cite{acta09}  for the case of two-gluon emission.

In     Fig.~\ref{fig:colcoh}a       two soft  gluons    
 with   momenta $q_1$ and $q_2$  are  produced  from a fast parton 
 with   momentum $p$. Suppose 
$q_2^0 \ll   q_1^0$. 
We    distinguish two  angular  regions for the softest gluon $q_2$~\cite{acta09}. 
i)~When  $q_2$ is  at small angle from $p$ ($q_1$),     then 
 the amplitude can be seen as the sequential emission of 
$q_1$ from $p$ and of $q_2$ from $p$ ($q_1$). This corresponds to the 
standard  bremsstrahlung picture 
 based on  radiation cones centered around $p$ and $q_1$.  
ii) When  $q_2$ is  at large angle, $\theta_{p q_2} \gg \theta_{p q_1}$, 
then the  directions of $p$ and $q_1$ can be identified and the two 
 emission amplitudes act coherently to give (Fig.~\ref{fig:colcoh}b) 
what  can be seen  as the sequential emission of 
$q_2$ from $p$ and of $q_1$ from $p$. 
The reversed order of the emissions compared to case i) 
 reflects the fact that the radiated gluon sees  the 
total color charge of the emitting jet. 
Fig.~\ref{fig:colcoh}b    illustrates  that  contributions 
of different emitters combine  to give an effective contribution 
in which the emissions are ordered in angle, so that 
angular ordering~\cite{marchweb80s,dok88,mc89} 
   replaces energy ordering.

The  above framework  of  collinear  
showering     supplemented with  phase space   constraints designed 
to implement  the  angular ordering~\cite{mc_lectures}  
forms the  
basis  of standard shower Monte Carlo event generators.   
We next discuss the modifications of this framework  that  are required 
to treat   parton showers at  increasingly high energies.

\subsection{Space-like parton  shower  at high energies} 

New effects arise  if one is to  extend   
 the picture of the above subsection     
 to  the case of   
very  high  energies, where processes   with multiple  hard scales 
become significant.   
The first  new effect  
  is that 
    soft-gluon     insertion rules~\cite{bcm,dok88}     
  for  $n$-parton scattering    amplitudes  $M^{(n)}$  
  can still be given in terms of   real and virtual  soft-gluon 
currents ${\bf J}^{(R)}$ and   ${\bf J}^{(V)}$, 
\begin{eqnarray}
\label{softinsert}
{| M^{(n+1)} (k , p) |}^2 &=&  \, 
\left\{  
[ M^{(n)}(k+q, p) ]^{\dagger} \; 
[{\bf J}^{(R)}]^2  \; 
{ M^{(n)}  } (k+q , p) \right.
\nonumber\\
&-& \left.
[ M^{(n)}(k, p) ]^{\dagger} \; 
[{\bf J}^{(V)}]^2 \; 
{ M^{(n)}  } (k , p) 
\right\}   \;\;  ,  
\end{eqnarray}
 but   the  currents are 
 modified in the high-energy, multi-scale region by terms that 
depend on the total transverse momentum transmitted down the 
initial-state parton decay chain~\cite{hef90,skewang,mw92}. 
For this reason  the physically relevant 
distribution to describe space-like  showers   at high energies    
is    not 
 an  ordinary parton density but rather  a  TMD  parton density.

The next  point   concerns the structure of virtual 
corrections.  Besides   Sudakov form-factor  contributions  included in    standard  
shower algorithms~\cite{mc_lectures,eswbook},  one needs  in general 
virtual-graph terms to be incorporated in 
transverse-momentum dependent (but universal)   splitting 
functions~\cite{skewang}.  These     allow  one    
 to take  account of    gluon coherence  not only for 
collinear-ordered emissions but also in the non-ordered region that 
opens up at high $\sqrt{s} / p_\perp$,  where $\sqrt{s}$ is the 
total center-of-mass energy and $p_\perp$  is  
the  typical   transverse momentum of a  produced jet.

\begin{figure}[htb]
\vspace{36mm}
\includegraphics{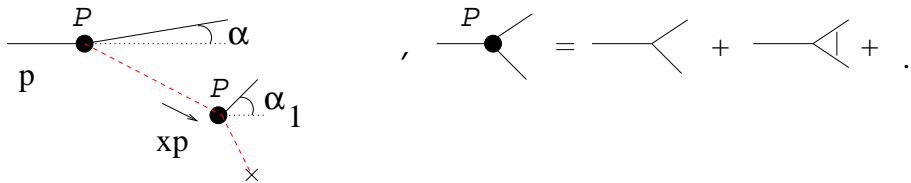}
\caption{\it   \small   (left) Coherent radiation 
 in the space-like parton shower for $x \ll 1$; (right) the unintegrated 
splitting function ${\cal P}$, including small-$x$ virtual 
corrections.} 
\label{fig:coh}
\end{figure}

   The  resulting  structure of the  parton branching, depicted in 
  Fig.~\ref{fig:coh},  
    differs  from that   in  
   Eq.~(\ref{psbasic}):  
  the branching probability is k$_\perp$-dependent, and   part 
    of the virtual  corrections are  associated to the (unintegrated) 
    splitting functions.      Schematically one  has,  
    using the recursion relation (\ref{softinsert}),     
\begin{eqnarray}
\label{uglurepr1}
  {\cal G} ( x , \kt , \mu ) & = & 
  {\cal G}_0 ( x , \kt , \mu ) + 
\int { {dz} \over z} \int { { d q^2} \over q^2} \ 
\Theta   (\mu - z  q) \
\nonumber\\
& \times & 
 \Delta    (\mu , z  q) 
\ {\cal P} ( z, q, \kt)   
\   {\cal G} 
 ( { x \over z} , \kt  + (1-z) q, q )
 \hspace*{0.3 cm} ,         
\end{eqnarray} 
where $ {\cal G}  $ is the unintegrated gluon distribution, $\Delta $ is the 
form factor, and 
${\cal P}$ is the unintegrated splitting function (Fig.~\ref{fig:coh}).      
   The  kernels 
 ${\cal P}$ depend  on 
transverse momenta and include part of the virtual corrections, in   such a  way  as 
  to avoid double  counting with the Sudakov form factor  $\Delta$, while 
 reconstructing  color coherence not only at large 
 $x $ but also at small $x$   in the angular region (Fig.~\ref{fig:coh}) 
\begin{equation}
\label{cohregion}
\alpha / x  > \alpha_1 > \alpha \hspace*{0.3 cm}    , 
\end{equation} 
where the angles  $\alpha$ for the 
partons radiated from the initial-state 
shower are taken with respect to the 
initial beam jet direction, and increase with increasing off-shellness.

In terms  of  high-energy  power-counting,    the   effects   of   region   
(\ref{cohregion})  are potentially enhanced by terms 
\begin{equation}
\label{superlead}
\alpha_s^k \ln^{k + m}  \sqrt{s} / p_\perp   
 \hspace*{0.3 cm}    . 
\end{equation} 
In inclusive processes,  coherence leads to strong cancellations 
between reals 
 and virtuals  so that terms 
 with $m  \geq 1$  in  Eq.~(\ref{superlead})   drop out, 
 and high-energy corrections  are  at most   
 single-logarithmic~\cite{ch94,jaro,fadlip98}.  
For exclusive  jet  distributions  such cancellations  are not  present 
and one may expect   stronger enhancements. 
 The   implementation 
of coherent-branching effects   associated with  high-energy logarithms   
is, from the point of view  of jet physics, 
 the main motivation for developing the formalism 
   of unintegrated parton distributions 
and  implementing it  in shower Monte Carlo event generators. 
See~\cite{hj_ang}  for 
recent phenomenological 
investigations of such effects on the structure of angular correlations and 
multiplicity distributions  in multi-jet final states, 
and~\cite{deak_etal_higgs} for applications to the jet structure associated with 
heavy mass states.

Monte Carlo implementations of  the modified branching described in this 
section  include  the  parton-shower  event generators 
\cascade~\cite{casc}, \ldcmc~\cite{ldc}.   
See~\cite{hj_rec,acta09,ajalt}  for further references. 
These implementations 
take into account   gluon coherence effects 
and include TMD gluon distributions. 
An extension to  valence quark distributions is given  in~\cite{epr-1012}.   
In the next section we describe recent  work to go  beyond this 
approximation and include sea quark contributions.

\section{Sea quark distribution and Drell Yan production}

\begin{figure}[htb]
\vspace{42mm}
\includegraphics{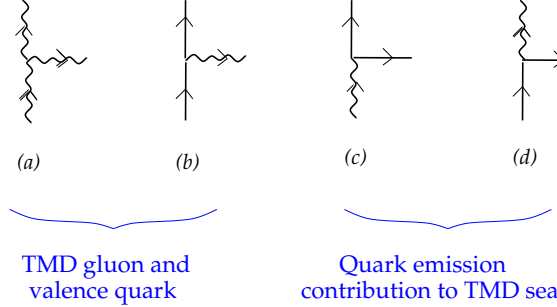}
\caption{\it  \small      Splitting vertices for TMD pdf evolution.}
\label{fig:splits}
\end{figure}

Monte Carlo calculations based on TMD approaches have so far 
been carried out mostly in a  quenched approximation in which 
only gluon and valence quarks are included. 
In terms of  the parton branching kernels, 
this  corresponds to taking into account  only contributions from the 
 splitting vertices {(a)}  and   {(b)}   in Fig.~\ref{fig:splits}.  In this section 
we briefly describe the work~\cite{hent11}      to  go  beyond this 
quenched approximation and treat  TMD  sea quark contributions. 
  The main focus of this   work  is to  take into account  effects of   
 the     splitting process     in Fig.~\ref{fig:splits}{(c)}     at TMD level.

Let us  recall that early  attempts~\cite{kwie03,wattetal03}  to 
 treat the unintegrated pdf evolution 
  beyond the quenched approximation  include  quarks 
    via   splitting probabilities to lowest order of perturbation 
  theory,     neglecting  
   any transverse momentum  dependence in the branching.  
In~\cite{wattetal09}  k$_\perp$-dependent 
kinematic corrections are   included, while the  
 splitting  kernels  are still taken in lowest order.

Also,  a program to  perform shower Monte Carlo evolution 
at unintegrated level has recently been proposed~\cite{jadach09}   
based on the expansion~\cite{CFP} in 
two-particle irreducible (2PI)  kernels.  This program is formulated  
 at the next-to-leading order. However,  this 
is limited at present  to the flavor nonsinglet sector,  i.e. 
does not include  small-$x$  logarithmic  corrections    which  are present 
beyond  NLO  in  flavor singlet   distributions.

\begin{figure}[htb]
\vspace{72mm}
\includegraphics{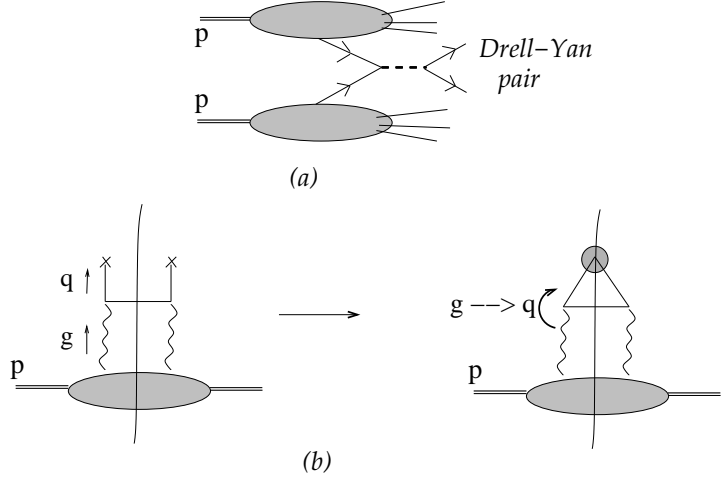}
\caption{\it   \small   (a) $ {\overline q}  q $  Drell-Yan production; (b) $ g \to q $ splitting 
contribution to sea quark   distribution.} 
\label{fig:qg-split}
\end{figure}

The approach  taken  in~\cite{hent11}    
is  based on    the high-energy form of the 
(off-shell)  TMD  quark Green function defined in~\cite{ch94}. This    
is obtained by    generalizing  to finite transverse momenta, in the high-energy 
region,   the 2PI expansion~\cite{CFP}.  The  k$_\perp$-dependent  gluon-to-quark 
 splitting        kernel    is depicted  in  Fig.~\ref{fig:qg-split}{(b)} .   
Thus, unlike the approaches in~\cite{kwie03,wattetal03,wattetal09}, on one hand, 
and in~\cite{jadach09}, on the other hand, 
     Ref.~\cite{hent11} constructs  TMD quark distributions  
by using  the k$_\perp$-dependent  gluon-to-quark 
 splitting function,   which governs 
 sea quark evolution to   all orders in perturbation theory 
 in the small-$x$ limit.  
  Refs.~\cite{hent11,hent-prep}  investigate 
 the phenomenological relevance of the finite $  k_\perp $  terms 
 for Drell-Yan production (Fig.~\ref{fig:qg-split}{(a)}). 

Schematically, the  TMD gluon-to-quark kernel   has the form 
\begin{equation} 
\label{pqg}
{\cal P}_{g \to q}  ( z;  q_\perp,  k_\perp )  =    P_{qg}^{(0)}  ( z) \ 
\left( 1 +  \sum_{n=1}^{\infty}   \ b_n (z )   ( k_\perp^2 /  q_\perp^2 )^n  \right)  \;\;\; , 
\end{equation}  
where $P^{(0)} $  is  the lowest-order  DGLAP  splitting function, 
$z$ is the longitudinal momentum transfer, 
$ k_\perp$ and $  q_\perp $ are respectively the gluon and quark  
transverse momenta in Fig.~\ref{fig:qg-split}{ (b)}, 
and all coefficients $b_n (z) $ of the  finite terms in  $  k_\perp $
  are known from~\cite{ch94}.  Although  it is evaluated off-shell,  the 
  splitting probability in Eq.~(\ref{pqg}) is  
   universal~\cite{ch94,cc05}.   The approaches  
 in~\cite{kwie03} and~\cite{wattetal03,wattetal09}  only  include the 
 term $P^{(0)} $   in Eq.~(\ref{pqg}), while Ref.~\cite{hent11}  includes the full series. 
 The finite terms in  $  k_\perp $ are  the ones  responsible for determining   
  small-$x$  logarithmic corrections  
to  flavor-singlet  quark evolution   to all orders in $\alpha_s$.

In Ref.~\cite{hent-dis}    contributions from 
the  TMD flavor-singlet quark  distribution   are implemented 
in the parton shower Monte Carlo 
event generator \cascade~\cite{casc}.  
This  constitutes a starting point  to systematically include quark emissions 
in the parton shower at TMD level.   
At present  the  implementation  is done in such a way that    
  the shower couples to quarks  only once, while  a  complete 
 implementation will allow  this to occur arbitrarily many times. 
 Nevertheless,  already   the present 
  implementation incorporates  for the first time   the small-$x$ 
    dynamics   encoded in Eq.~(\ref{pqg}) in a  parton shower, and   makes it 
    possible to treat in this framework      hard 
processes  induced by sea quarks  
on the same footing as processes induced by gluons.

\begin{figure}[htb]
\vspace{36mm}
\includegraphics{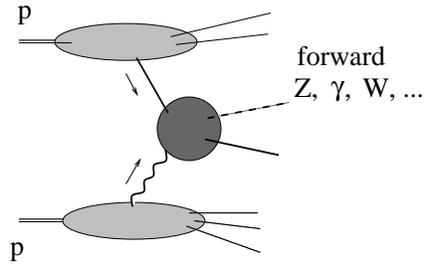}
\caption{\it   \small   Forward Drell-Yan production.}
\label{fig:fwdDY}
\end{figure}

 Refs.~\cite{hent11,hent-dis}  present     results  within this framework 
 for Drell Yan production processes.    This    is 
phenomenologically relevant,  because     
   Drell Yan  production  at the LHC    (Fig.~\ref{fig:qg-split}{\em (a)})    
 receives large contributions from 
  sea quark scattering at small $x$~\cite{ajalt}.   Such  contributions 
 affect  many aspects of  LHC physics, as 
Drell-Yan   processes are  instrumental  in  precision electroweak  measurements, 
in luminosity monitoring  and pdf determinations,  and in new physics searches. 
Results 
    on high-energy Drell-Yan at TMD level have so far been obtained in 
    the  $ q g^* $    channel~\cite{marball}, and in the associated production  channel 
     $Z / W  + $ heavy quarks~\cite{michal08}. 
    The   $ q g^*$   channel is   of  direct phenomenological 
       relevance for forward Drell-Yan production.   
 Ref.~\cite{hent11}      evaluates the perturbative  coefficients  for  the 
 coupling of the  TMD  sea quark  distribution    to Drell-Yan  
 by using the ``reggeized quark"  calculus~\cite{lip-q}. 
This   extends 
 the high-energy effective action formalism~\cite{Lipatov:1995pn},  
 currently explored at NLO~\cite{Hentschinski:2011tz}, to  
amplitudes with 
 quark $t$-channel exchange~\cite{Fadin:1976nw,knie-sale}. 
  In~\cite{hent11,hent-dis}  the    
method   is employed to  investigate   predictions for   
  Drell-Yan  production in the forward region  (Fig.~\ref{fig:fwdDY}). 

\begin{figure}[htb]
\vspace{56mm}
\includegraphics{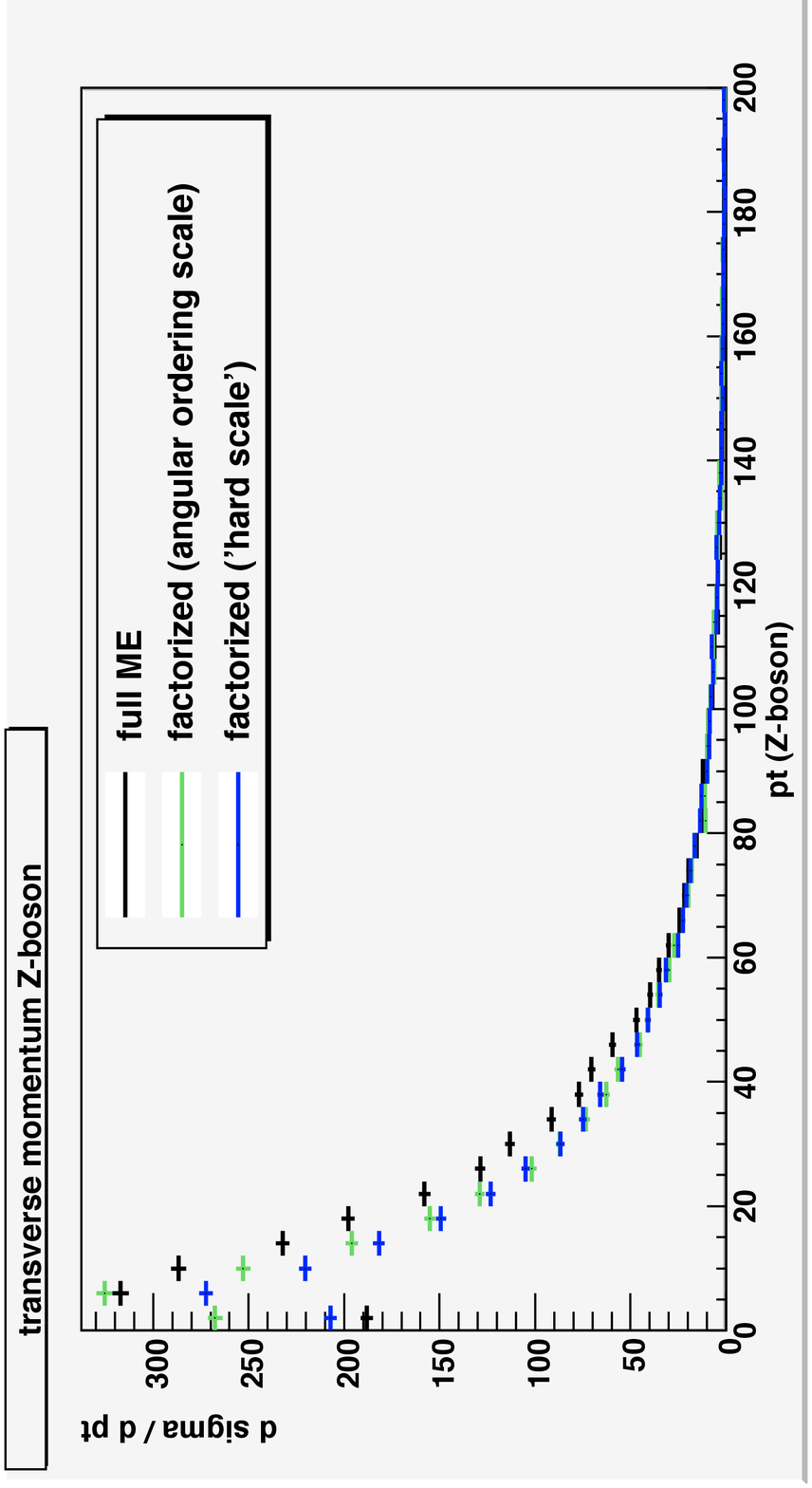}
\includegraphics{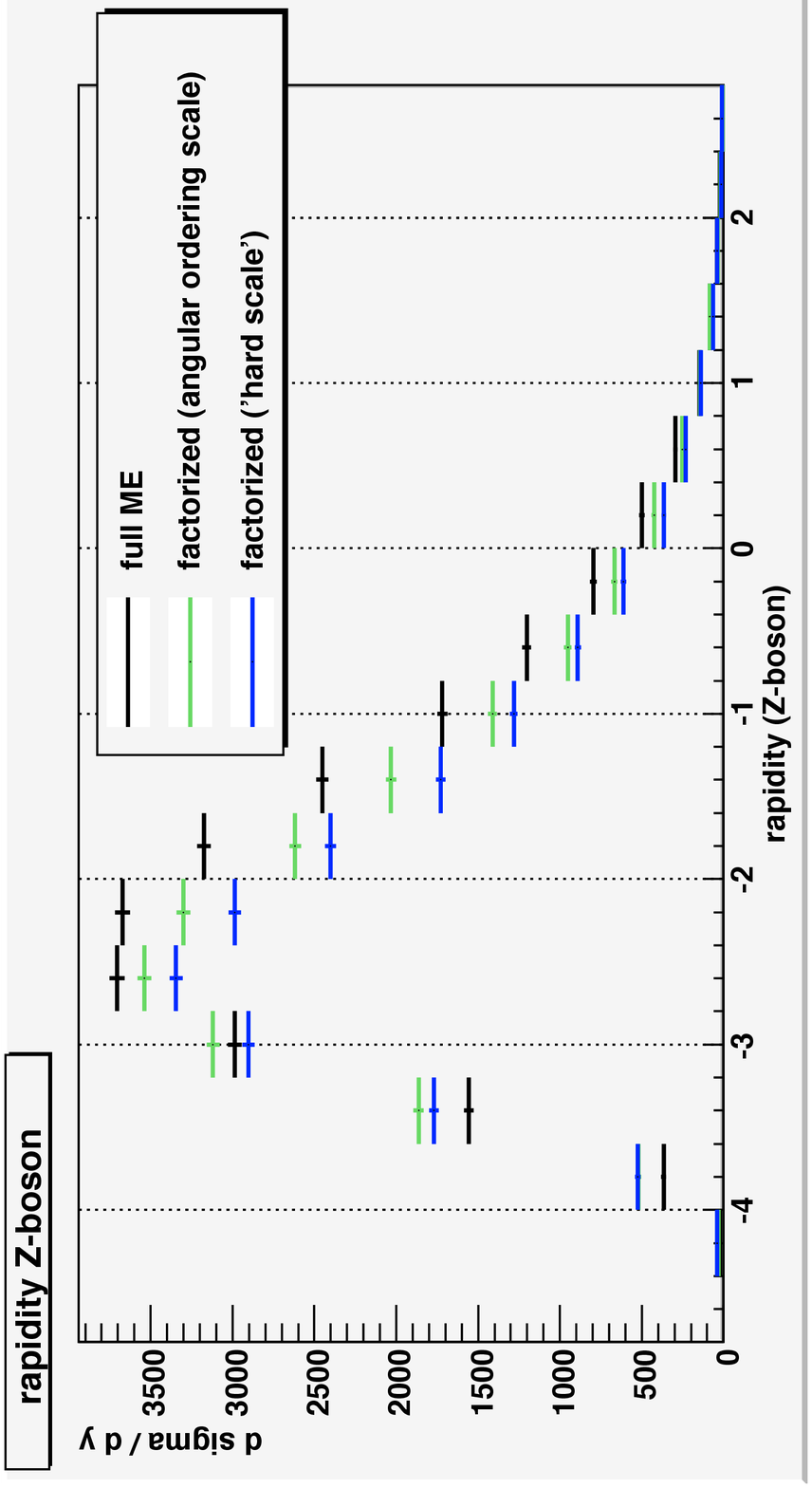}
\caption{\it   \small  Z-boson  transverse-momentum and rapidity spectra.}
\label{fig:Zspectra}
\end{figure}

\begin{figure}[htb]
\vspace{56mm}
\includegraphics{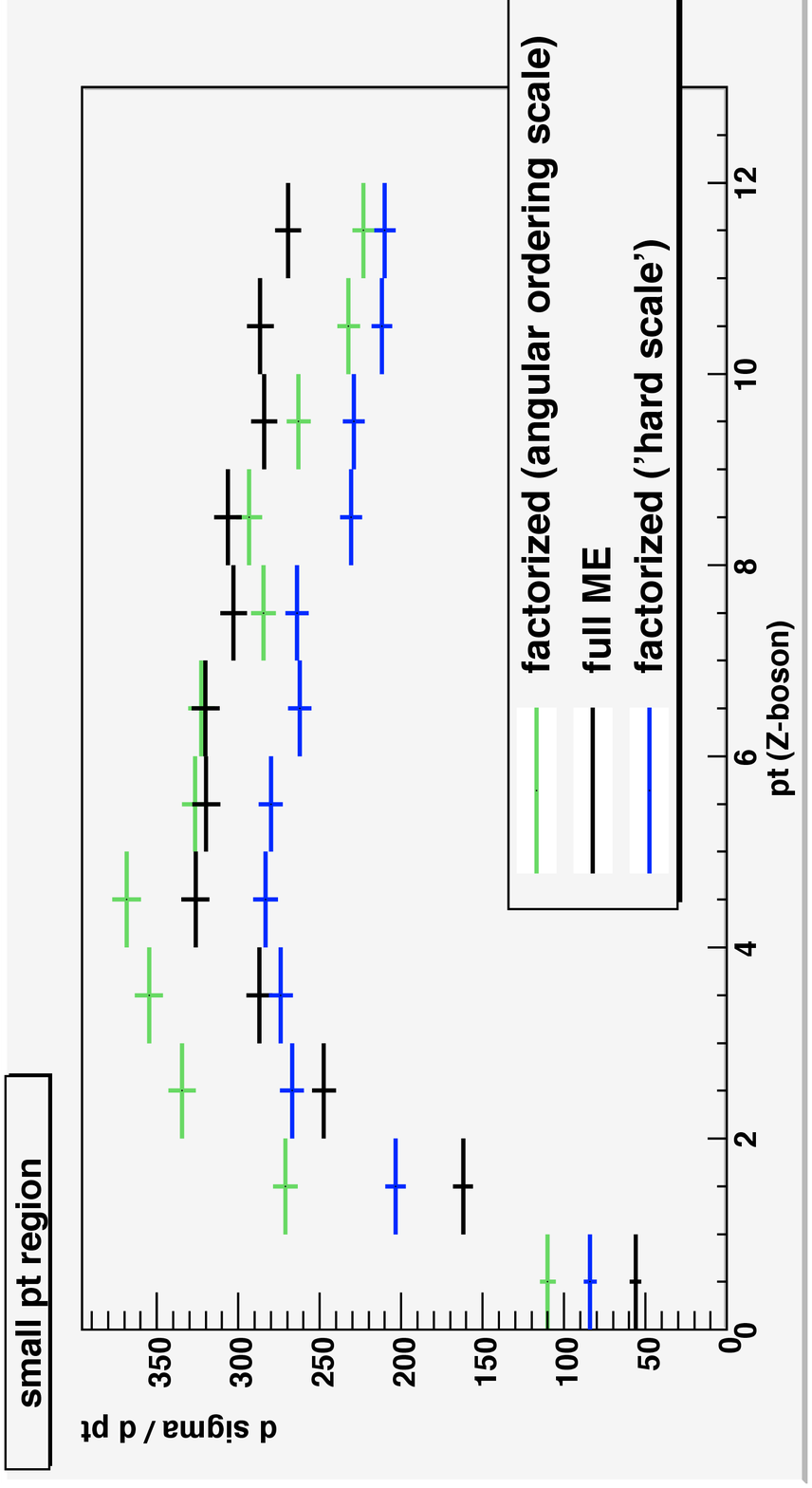}
\includegraphics{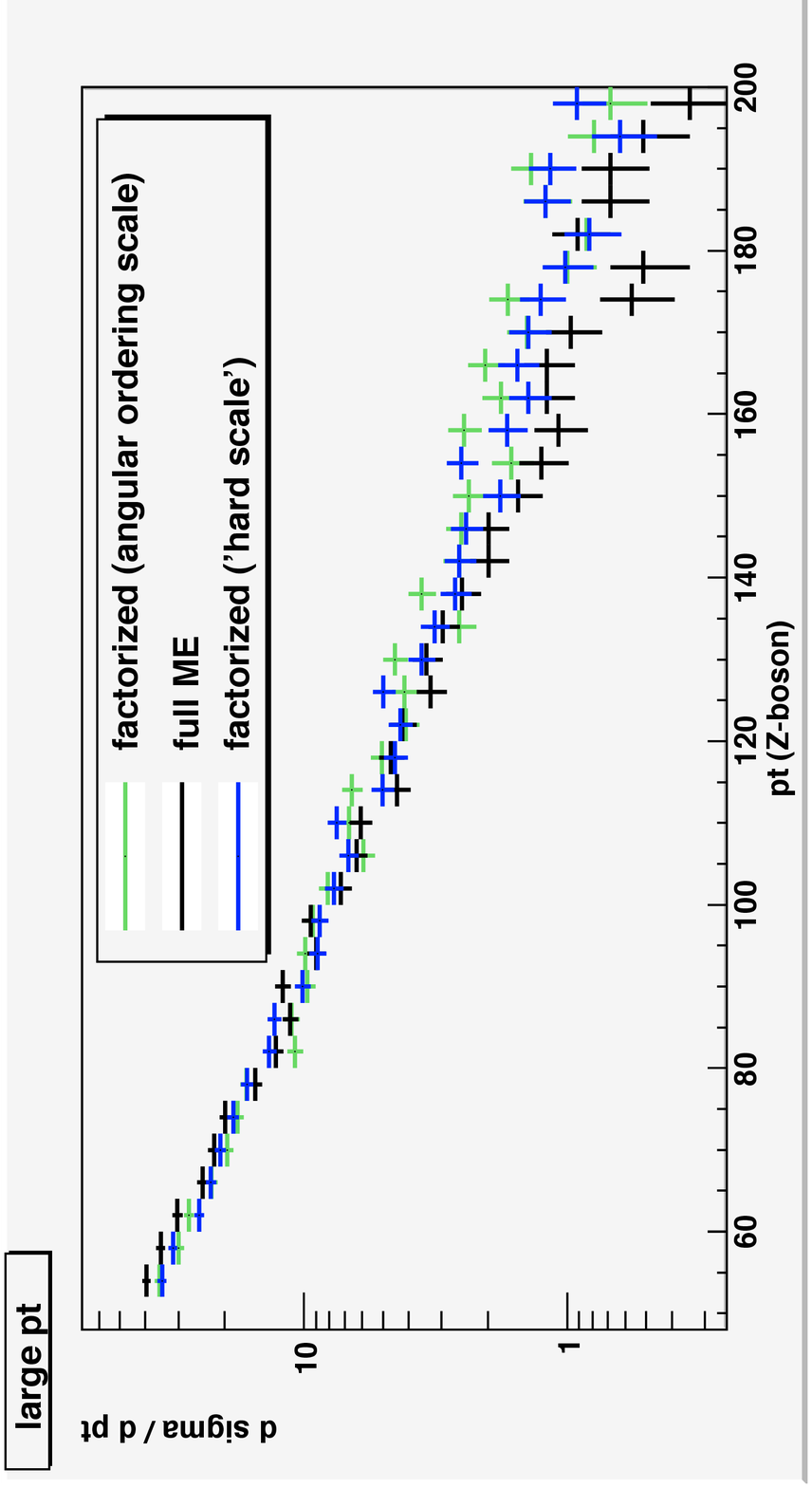}
\caption{\it  \small    Numerical comparison 
 in the small-p$_T$  and large-p$_T$ regions.}
\label{fig:Zcompar}
\end{figure}

In Figs.~\ref{fig:Zspectra},\ref{fig:Zcompar}  we  report 
numerical results  for $Z$-boson rapidity and transverse momentum spectra. 
We present  results for two  possible choices of the   scale  $\mu$  in the  
shower  evolution equation (\ref{uglurepr1}): one in which $\mu$ is set 
equal to the hard scale, defined by    
\begin{equation} 
\label{hard-scale} 
\mu^2  =   p_\perp^2 + M_Z^2 \;\; ,  
\end{equation} 
where $p_\perp$ and $M_Z$ are  the transverse momentum and mass of the 
$Z$ boson; another in which $\mu$ is set 
equal to the maximum-angle scale determined by the angular ordering kinematics, 
given by 
\begin{equation} 
\label{angord-scale} 
\mu^2  =  {{   q_\perp^2 + (1-z)  k_\perp^2 } \over  {  (1 - z )^2 }}  \;\; ,  
\end{equation} 
where the variables are as specified  below Eq.~(\ref{pqg}).   
Since  we are working in the forward region, 
we  also include results  obtained 
from      the  $ q g^* $ matrix element.

All three curves in Figs.~\ref{fig:Zspectra},\ref{fig:Zcompar}  correspond to 
calculations which  are  applicable at forward rapidity.   The  shape  of 
the  rapidity spectrum in Fig.~\ref{fig:Zspectra}, falling off for central 
production,   can thus be regarded as defining  the kinematic region of applicability. 
The transverse momentum spectrum in Fig.~\ref{fig:Zspectra} 
indicates that the three calculations differ at low $p_t$, but they all converge 
for large  enough  $p_t$.  This is illustrated in more detail in 
Fig.~\ref{fig:Zcompar}.  The difference  
between the matrix element and  factorized calculations 
in the small-$p_t$ region is mostly due to  quark $s$-channel contributions. 
This region is dominated by Sudakov form factor effects,  giving the 
turn-over in the $p_t$ spectrum, which are sensitive to the different 
choices of scales in Eqs.~(\ref{hard-scale}),(\ref{angord-scale}).  
The difference between the curves 
 disappears  in the large $p_t$ region as  the 
quark $t$-channel  contribution dominates, and   contributions from the parton showers 
also become relatively  insensitive to the different scale choices, because transverse 
momentum ordering sets in driven by the high $p_t$.

The numerical results in  Figs.~\ref{fig:Zspectra},\ref{fig:Zcompar} 
 can  be  seen as a consistency check  
on the   approach  based on TMD parton branching 
as regards forward Drell-Yan production. 
 The result~\cite{hent11} for the unintegrated flavor-singlet quark  distribution 
is however more general,  and is a  necessary ingredient to  develop  a  description of 
   vector boson  production  and associated final  states 
   throughout    the rapidity range of     forward  and central   regions. 
This is the subject of  forthcoming   work.

\section{Summary}

We have discussed recent  progress on 
transverse momentum dependent parton distributions. Relying on 
TMD factorization results at small $x$, we have discussed applications to 
branching Monte Carlo methods  for  initial-state  parton showers 
in high-energy  collisions.  Calculations based on these methods have mostly 
been developed within a quenched approximation in which only gluon and valence 
quark effects are taken into account at  
unintegrated  level. We have reported work to 
go  beyond this approximation by including sea quark  contributions. 

The method is based on taking into account 
 flavor-singlet quark evolution 
via the  transverse momentum dependent 
  gluon-to-quark  splitting kernel in  Eq.~(\ref{pqg}). 
We have  used  this  along  with  the  
 Monte Carlo  implementation~\cite{casc} 
of the  small-$x$ coherent branching  equation given  in Eq.~(\ref{uglurepr1}). 
Compared to previous  approaches to treat the TMD sea quark   
distribution,     the main feature of this approach   
  is that it includes, in addition to the 
  lowest-order  splitting function,      
    the full  series of    finite-$  k_\perp $  terms  in the 
  gluon-to-quark  two-particle irreducible kernel.  
  This allows one to sum small-$x$ logarithmic corrections to flavor-singlet 
 observables    to all orders in   $\alpha_s$.   

We have used this framework to obtain numerical predictions for 
forward  $Z$-boson production. To this end, 
  the perturbative  coefficients  for  
 coupling  the  TMD  sea quark  distribution    to  Drell-Yan  production  
 have been determined  by using the  formalism~\cite{lip-q}
for high-energy quark $t$-channel exchange.    
We have  investigated  the dependence  of the results 
on the shower evolution scale.  

The method proposed in this  work   
is implemented at present by letting quarks interact with the shower only once. 
Using the ingredients discussed in this work, 
this could however   be   extended  by constructing  
a  fully  coupled shower evolution in the flavor-singlet sector. 
Also,  only one  coupling of the sea quark  to  vector bosons  
is evaluated at TMD level at present. This should  be extended 
in order  to  treat  Drell-Yan  throughout  
  the rapidity range of     forward  and central   regions.
Nevertheless, the results presented  here, being   the first implementation 
of small-$x$  sea-quark dynamics in a parton shower, constitute a 
starting point to include quark emissions  systematically and to treat 
hard processes initiated by sea quarks on the same footing as  processes 
initiated by gluons.

\vskip 1 cm 

\noindent 
{\bf Acknowledgments}. 
We thank  the  organizers   for the   kind   invitation and  opportunity to present this 
work at the conference.  We   gratefully acknowledge the 
pleasant  atmosphere  and fruitful discussions   at the meeting.

\begin{footnotesize}

\end{footnotesize}

\end{document}